\documentclass[aps,twocolumn, amsmath,amssymb, superscriptaddress, reprint]{revtex4-1}
\usepackage{hyperref}
\usepackage{amsmath,amsthm,amssymb}
\usepackage{xcolor}
\usepackage{graphicx}
\usepackage{textcomp}
\usepackage{siunitx}

\usepackage{tikz}

\usepackage{pgfplots}
\renewcommand{\vec}[1]{\boldsymbol{#1}}
\newcommand{\dx}{\,\text{d}\vec{x}}

\begin{document}

\title{Solving Large-Scale Inverse Magnetostatic Problems using the Adjoint Method}

\author{Florian Bruckner}
\thanks{Correspondence to: \href{mailto:florian.bruckner@tuwien.ac.at}{florian.bruckner@tuwien.ac.at}}
\affiliation{Christian Doppler Laboratory of Advanced Magnetic Sensing and Materials, Institute of Solid State Physics, Vienna University of Technology, Austria}
\author{Claas Abert}
\affiliation{Christian Doppler Laboratory of Advanced Magnetic Sensing and Materials, Institute of Solid State Physics, Vienna University of Technology, Austria}
\author{Gregor Wautischer}
\affiliation{Christian Doppler Laboratory of Advanced Magnetic Sensing and Materials, Institute of Solid State Physics, Vienna University of Technology, Austria}
\author{Christian Huber}
\affiliation{Christian Doppler Laboratory of Advanced Magnetic Sensing and Materials, Institute of Solid State Physics, Vienna University of Technology, Austria}
\author{Christoph Vogler}
\affiliation{Institute of Solid State Physics, Vienna University of Technology, Austria}
\author{Michael Hinze}
\affiliation{Department of Mathematics, University of Hamburg, Germany}
\author{Dieter Suess}
\affiliation{Christian Doppler Laboratory of Advanced Magnetic Sensing and Materials, Institute of Solid State Physics, Vienna University of Technology, Austria}

\begin{abstract}
An efficient algorithm for the reconstruction of the magnetization state within magnetic components is presented. The occurring inverse magnetostatic problem is solved by means of an adjoint approach, based on the Fredkin-Koehler method for the solution of the forward problem. Due to the use of hybrid FEM-BEM coupling combined with matrix compression techniques the resulting algorithm is well suited for large-scale problems. Furthermore the reconstruction of the magnetization state within a permanent magnet is demonstrated.
\end{abstract}

\maketitle

\section{Introduction}
Magnetic materials are used in a wide range of applications, ranging from permanent magnets, magnetic machines, up to magnetic sensors and magnetic recording devices. Solving the Inverse Magnetostatic Problem allows to reconstruct the internal magnetization state of a magnetic component, by means of magnetic field measurements outside of the magnetic part, which is of importance for quality control. Compared with the forward problem, where the magnetic state is known and the strayfield is calculated, inverse problems are much harder to solve, since they typically are much worse conditioned and often not uniquely solveable. Inverse problem solvers are based on stable and reliable solvers for the forward problem. In the case of magnetostatic Maxwell equations, Finite Element (FEM) formulations, combined with methods to handle the open-boundary, have proven to be the methods of choice for many efficient and accurate methods \cite{brunotte_finite_1992, khebir_asymptotic_1990, fredkin_hybrid_1990}.

The applications of inverse problems can be coarsely divided into optimal design and source identification. Optimal design problems define a desired strayfield and try to calculate optimal material distributions or geometries to reach these requirements as accurate as possible \cite{neittaanmaki_inverse_1996, yoo_topology_2000}. In contrast to this, source identification problems try to reconstruct the state of existing magnetic components. The identification of (permanent) magnetic materials has e.g. been successfully applied for reconstructing the state of magnetic rollers used in copy machines \cite{igarashi_inverse_2000}, magnetically biased chokes \cite{husstedt_detailed_2014}, or even for the magnetic anomaly created by ferromagnetic ships \cite{chadebec_recent_2002}.

The presented solver for the inverse 3D magnetostatic Maxwell equations, is based on the well-established Fredkin-Koehler-Method \cite{fredkin_hybrid_1990}, which uses a hybrid FEM-BEM coupling for efficient handling of the open-boundary conditions. Combined with a hierarchical matrix compression technique for the dense boundary integral matrices, the algorithm is able to handle large-scale problems. Additionally, the use of a general Tikhonov regularization (see e.g. \cite{calvetti_tikhonov_2000}), provides a very flexible means to define application specific regularizations.

\section{Adjoint Method}
\subsection{Forward Problem}
The demagnetization field of a magnetic body is defined as $\vec{h}_\text{d} = - \vec{\nabla} u$, where the magnetic scalar potential $u$ is given by
\begin{alignat}{2}
  \Delta u &= \vec{\nabla} \cdot \vec{m} &&\quad\text{in}\quad \Omega \label{eq:demag_first} \\
  \Delta u &= 0                                        &&\quad\text{in}\quad \mathbb{R}^3 \setminus \Omega
\end{alignat}
with jump and boundary conditions
\begin{align}
  \left[ u \right]_{\partial \Omega} &= 0 \\
  \left[ \frac{\partial u}{\partial \vec{n}} \right]_{\partial \Omega} &= - \vec{m} \cdot \vec{n} \\
  u(\vec{r}) &\rightarrow \mathcal{O}(1/|\vec{r}|) \quad \text{for} \quad |\vec{r}| \rightarrow \infty
  \label{eq:demag_last}
\end{align}
where $\vec{m}$ is the magnetization and $\Omega$ is the magnetic region.

\begin{figure}
  \centering
  \begin{tikzpicture}
    \node[anchor=south west,inner sep=0] (image) at (0,0) {
      \includegraphics[width=0.7\columnwidth]{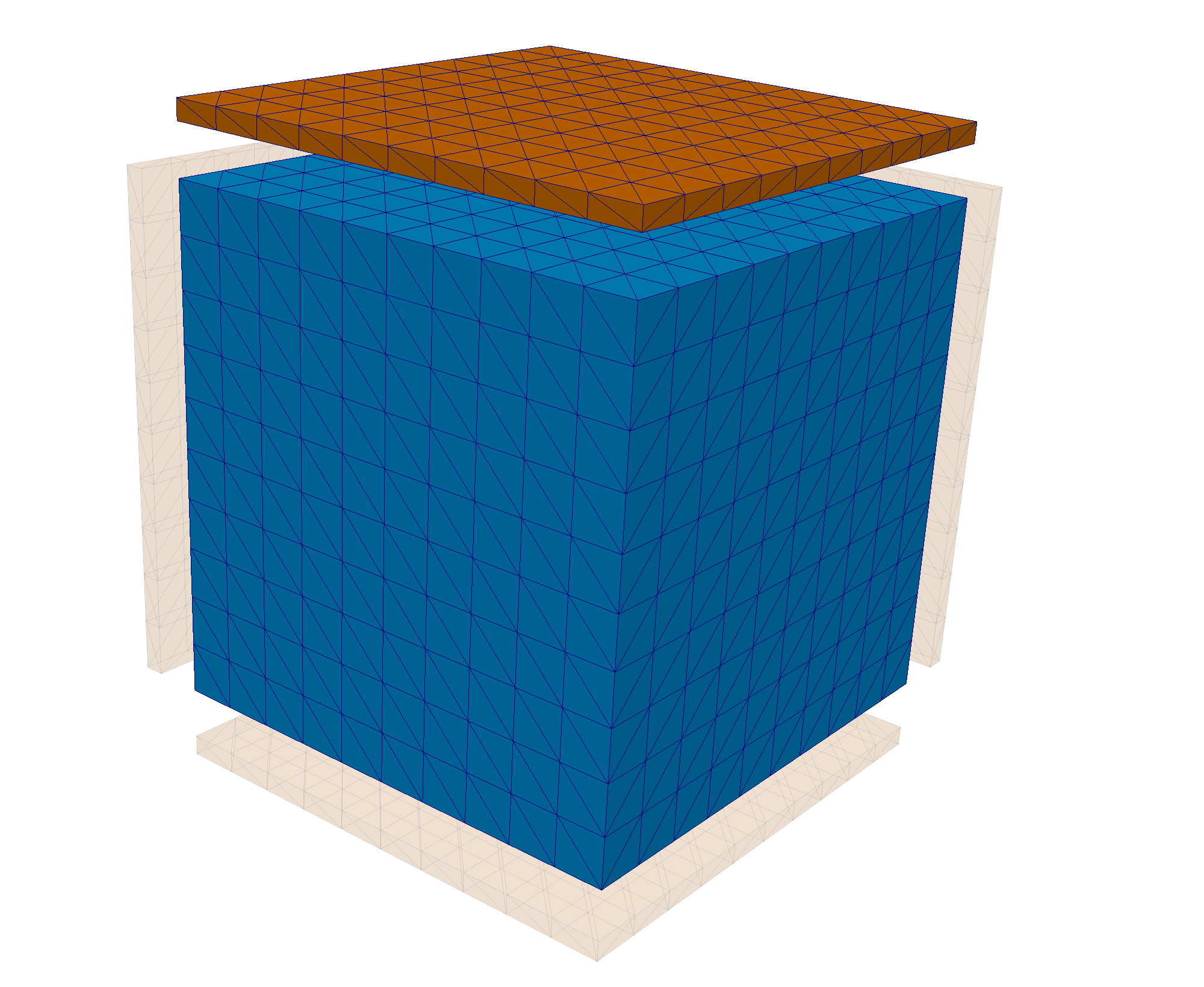}
    };
    \begin{scope}[x={(image.south east)},y={(image.north west)}]
      \draw[fill] (0.70, 0.87) circle[radius=0.005, fill] -- (0.80, 0.92) node[right] {$\Omega_\text{h}$};
      \draw[fill] (0.70, 0.40) circle[radius=0.005, fill] -- (0.80, 0.45) node[right] {$\Omega_\text{m}$};
    \end{scope}
  \end{tikzpicture}
  \caption{
    Discretized magnetization region $\Omega_\text{m}$ (blue) and measurement region $\Omega_\text{h}$ (brown). Since the strayfield problem is scale-invariant, length units are omitted. The magnetization is defined on a unit cube. Measurement boxes of thickness $0.04$ are located next to each side of the cube, using an airgap of $0.1$ (the 2 boxes in front are not shown in the figure).
  }
  \label{fig:mesh}
\end{figure}
The forward problem requires the solution of the potential $u$ on the region $\Omega_\text{h}$ generated by the magnetization in a magnetic region $\Omega_\text{m}$ (see Fig. \ref{fig:mesh}).
This problem is solved by considering a single region $\Omega = \Omega_\text{m} \cup \Omega_\text{h}$ with $\vec{m}(\vec{x}) = 0$ for $\vec{x} \in \Omega_\text{h}$.
The hybrid FEM/BEM method introduced by Fredkin and Koehler \cite{fredkin_hybrid_1990} is one of the most accurate methods for the solution of this problem and will be used in the following.
Consider the following splitting of the solution $u$:
\begin{equation}
  u = u_1 + u_2
\end{equation}
Here $u_1$ is defined by
\begin{alignat}{2}
  \Delta u_1 &= \vec{\nabla} \cdot \vec{m} && \quad \text{in} \quad \Omega \\
  \frac{\partial u_1}{\partial \vec{n}} &= - \vec{m} \cdot \vec{n} && \quad \text{on} \quad \partial\Omega \\
   u_1 &= 0 && \quad \text{in} \quad \mathbb{R}^3 \setminus \Omega.
\end{alignat}
This Neumann problem is solved with the finite-element method.
While $u_1$ solves for the right-hand side $\vec{m}$ and fulfills the jump condition of the normal derivative $-\vec{m} \cdot \vec{n}$, it is not continuous across $\partial \Omega$ as required.
This jump is compensated by $u_2$ which is defined as
\begin{alignat}{2}
  \Delta u_2 &= 0 && \quad \text{in} \quad \Omega \\
  [u_2] &= - [u_1] && \quad \text{on} \quad \partial \Omega \\
  \left[\frac{\partial u_2}{\partial \vec{n}}\right] &= 0 && \quad \text{on} \quad \partial\Omega \\
  u_2(\vec{x}) &= \mathcal{O}(1/|\vec{x}|) && \quad \text{if} \quad |\vec{x}| \rightarrow \infty
\end{alignat}
This system is solved by the following double-layer potential
\begin{equation}
  u_2 = \int_{\partial \Omega} u_1 \frac{\partial}{\partial \vec{n}} \frac{1}{|\vec{x} - \vec{x}'|} \dx
\end{equation}
For efficiency reasons the double-layer potential is only computed on the boundary $\partial \Omega$ using a Galerkin boundary-element method. Subsequently these values are used as Dirichlet boundary conditions to solve $u_2$ within $\Omega$ using the finite-element method.

All potentials are calculated using piecewise linear basis function ($\mathcal{P}_1$) and the derived strayfield would be constant within each element. Thus, a mass lumping procedure needs to be used to project the field onto piecewise linear basis functions which are defined on each vertex of the mesh.

\subsection{Inverse Problem}
The inverse problem can be understood as a PDE constrained optimization problem. Due to the ill-posedness of the inverse problem, some additional information needs to be provided to allow reasonable results. This can be achieved by using Tikhonov regularization, which uses an additional penalty term which should be considered for the optimization. A possible candidate for the objective function is
\begin{equation}
  J = \frac{1}{2} \int_{\Omega_\text{h}} \| -\vec{\nabla} u - \vec{h}_\text{m} \|^2 \dx
    + \alpha \, T(\vec{m})
\end{equation}
where $\vec{h}_\text{m}$ is the prescribed (measured) field in $\Omega_\text{h}$ and $\alpha$ is the Tikhonov constant corresponding to the regularization functional $T(\vec{m})$.
This functional should be minimized, constrained by
\begin{equation}
  F = \Delta u - \vec{\nabla} \cdot \vec{m} = 0 \quad \text{on} \quad \Omega
\end{equation}
with boundary conditions as given above. We aim to solve this problem using a gradient based iterative minimizer. The constraint $F$ gives an implicit expression for $u(\vec{m})$ which allows to directly calculate the desired gradient by
\begin{equation}
  \frac{\text{d} J}{\text{d} \vec{m}} = \frac{\partial J}{\partial u} \; \frac{\partial u}{\partial \vec{m}} + \frac{\partial J}{\partial \vec{m}}
\end{equation}
The inefficient calculation of the term $\frac{\partial u}{\partial \vec{m}}$ can be avoided using the adjoint approach, which makes use of the derivative of the constraint equation to eliminate the problematic term:
\begin{equation}
  \frac{\text{d} J}{\text{d} \vec{m}} = \lambda^T \frac{\partial F}{\partial \vec{m}} + \frac{\partial J}{\partial \vec{m}}
\end{equation}
where $\lambda$ is given by the adjoint equation
\begin{equation}
  \left(\frac{\partial F}{\partial u}\right)^T \lambda = - \left(\frac{\partial J}{\partial u}\right)^T.
  \label{eqn:adjoint}
\end{equation}
Since the Poisson problem is self-adjoint, the adjoint system \eqref{eqn:adjoint} can be solved along the lines of the forward problem.
Computing the variational derivative on the RHS yields
\begin{equation}
  \Delta \lambda = \vec{\nabla} \cdot (\vec{\nabla}u + \vec{h}_\text{m}) \quad \text{on} \quad \Omega.
  \label{eqn:demag_adjoint}
\end{equation}
where the sources (RHS) live on $\Omega_\text{h}$ and the solution is only computed on $\Omega_\text{m}$.
The same boundary conditions as for the forward problem hold.
Thus, the above described hybrid method is applied.
The gradient of $J$ is then finally given by
\begin{equation}
  \frac{\text{d} J}{\text{d} \vec{m}} = \vec{\nabla} \lambda + \alpha \, \frac{\partial T}{\partial \vec{m}}
  \label{eqn:demag_gradient}
\end{equation}
Note that $\vec{\nabla} u$ and $\vec{\nabla} \lambda$ are projected onto $\mathcal{P}_1$ before computing \eqref{eqn:demag_adjoint} and \eqref{eqn:demag_gradient} respectively.
The algorithm is implemented using Magnum.Fe \cite{abert_magnum.fe:_2013}, which is based on the finite element library FEniCS \cite{logg_automated_2012}. This allows a very comfortable definition of the regularization functional. Furthermore, automatic differentiation can be used for the calculation of the corresponding partial derivatives. The algorithm was verified by comparison with a FEM-only implementation, using the dolfin-adjoint library \cite{farrell_automated_2013}, which allows to automatically derive the adjoint equation for a given forward problem.

\section{Numerical Experiments}
The presented algorithm is validated by the reconstruction of the flower-state within a magnetic unit cube. The strayfield is calculated within measurement planes next to each side of the cube (see Fig. \ref{fig:mesh}). The magnetic state of the cube is parametrized by
\begin{align} \label{eqn:flower_state}
  \vec{m} = \begin{pmatrix} \sin(c_\text{tilt} \theta) \, \cos(\phi) \\ \sin(c_\text{tilt} \theta) \, \sin(\phi) \\ \cos(c_\text{tilt} \theta) \end{pmatrix} & &
  \begin{matrix} \theta=z \, \sqrt{x^2+y^2} \\ \phi=\tan^{-1}(y/x) \end{matrix}
\end{align}
where $c_\text{tilt}$ is an open parameter that allows to change the strength of the flower state. For the proper reconstruction of the magnetic state additional knowledge about the solution needs to be provided. For all presented results a smooth reconstructed magnetization is desired which suggests using the following default regularization functional
\begin{equation}
  T(\vec{m}) = \int_{\Omega_\text{m}} (\nabla \vec{m})^2 \dx
\end{equation}
For this specific flower-state the absolute value of the magnetization is known to be constant. Thus, the solution of the inverse problem could be simplified by using the following penaltization functional
\begin{equation} \label{eqn:constant_norm_regularization}
  T^*(\vec{m}) = \int_{\Omega_\text{m}} (\vec{m}^2-1)^2 \dx
\end{equation}
The assumption of a constant magnetization may be a good approximation for (isotropic) permanent magnetic materials. Due to the large magnetic remanence and the relatively small susceptibility the induced magnetization may be negligible (see \cite{bruckner_macroscopic_2016} for a simple model of isotropic permanent magnetic materials).

\begin{figure}
  \centering
  \begin{tikzpicture}
    \node[anchor=south west,inner sep=0] (image) at (0,0) {
      \includegraphics[width=0.7\columnwidth]{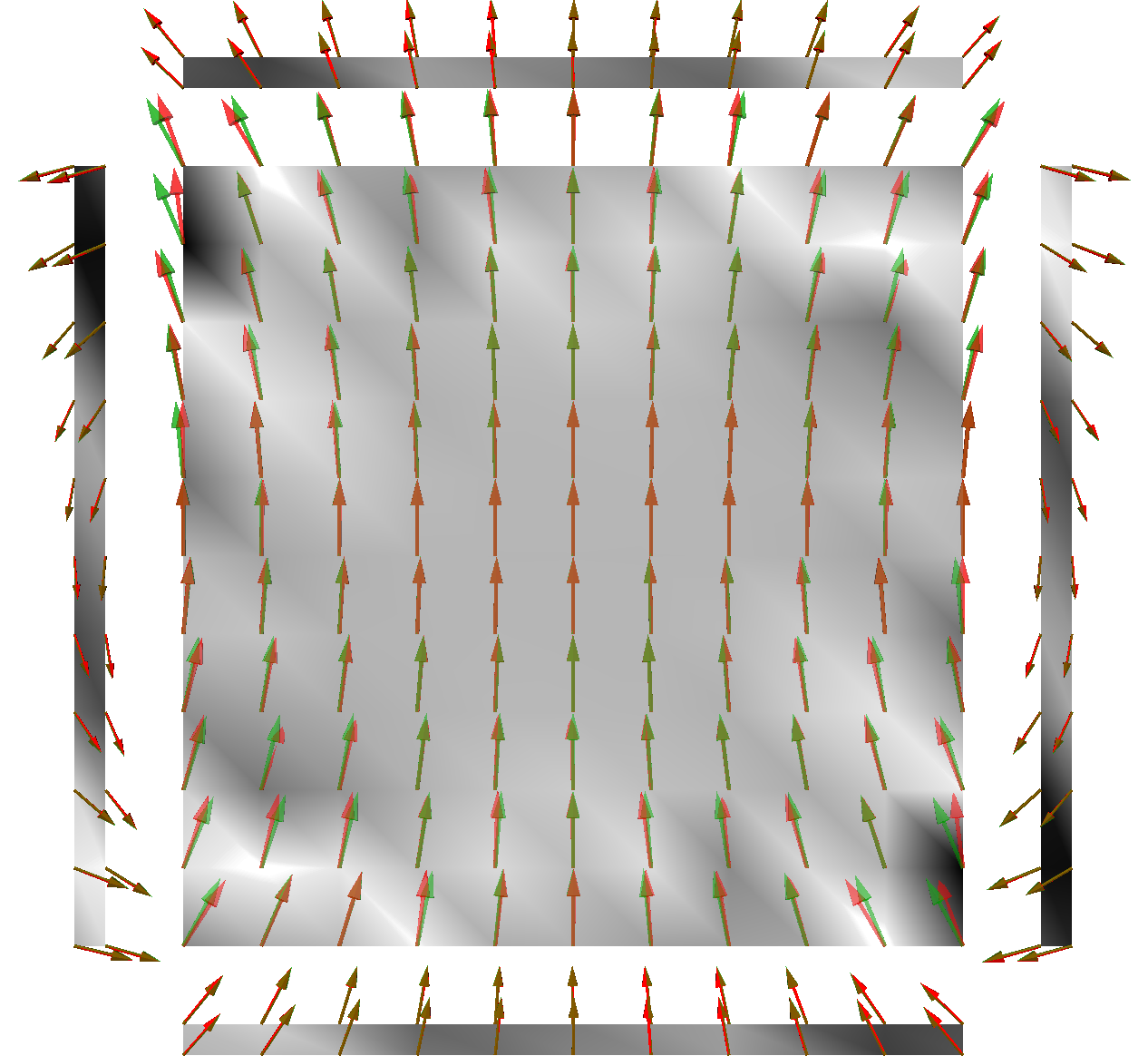}
    };
    \begin{scope}[x={(image.south east)},y={(image.north west)}]
      \draw[fill] (0.82, 0.93) circle[radius=0.005, fill] -- (0.84, 0.99) node[above] {$\Omega_\text{h}$};
      \draw[fill] (0.82, 0.82) circle[radius=0.005, fill] -- (0.89, 0.88) node[right] {$\Omega_\text{m}$};
    \end{scope}
  \end{tikzpicture}
  \caption{
    Reconstruction of a flower-state within a unit cube according to Eqn. \eqref{eqn:flower_state} using $c_\text{tilt}=2$. A cut through the $y=0$ plane is visualized. Starting from the initial parametrized flower state in $\Omega_m$ the magnetic strayfield is calculated within the fieldboxes $\Omega_h$ (green arrows). In order to simulate measurement errors a Gaussian noise with $\sigma=10^{-4}$ has been added to the forward strayfield. The reconstructed magnetization as well as strayfield are computed using $\alpha=10^{-3}$ (red arrows). The relative differences of initial and reconstructed states are indicated by the gray-scale. Maximal relative errors of the $x$-components amount to $0.25$ for the magnetization, and $5 \cdot 10^{-3}$ for the induced magnetic field, respectively.
  }
  \label{fig:diff}
\end{figure}

A Gaussian noise with zero mean and a standard deviation $\sigma = 10^{-4}$ has been added to the field, calculated by the forward problem, which should simulate unavoidable measurement errors. The minimization problem is solved by a gradient descent method combined with a line-search strategy. As expected, reconstruction without using a proper regularization leads to large magnetization vectors near to the edges of the unit cube. Increasing the regularization parameter $\alpha$, first avoids the over-fitting of the noisy measurement data, but finally leads to blurring of the reconstruction results if $\alpha$ gets too large. Determining the optimal alpha is a crucial step for the solution of an inverse problem. The results for the reconstruction of a flower state with $c_\text{tilt}=2$ using $\alpha=10^{-3}$ is visualized in Fig. \ref{fig:diff}. Although there are some deviations of the reconstructed magnetization from the reference state. It can be seen that the created strayfield is nearly identical. As stated above this is a clear indication of the bad condition of the inverse problem.

\begin{figure}
  \centering
  \begin{tikzpicture}
    \begin{axis}[
        legend pos=south west,
        legend style={font=\scriptsize},
        xlabel={Residual Norm $\| -\vec{\nabla} u - \vec{h}_\text{m} \|^2$},
        ylabel={Regularization Norm $\| \nabla \vec{m} \|^2$},
        xmode = log,
        ymode = log,
        ymin=1e-3, ymax=1e2,
        xtick={1e-4, 2e-4},
        xticklabels={$10^{-4}$,$2 \cdot 10^{-4}$},
        grid]
      \addplot[red, mark=+]   table[x index=2, y index=1] {6boxes_grad_None.dat};
      \addplot[green, mark=x] table[x index=2, y index=1] {6boxes_grad_1e-5.dat};
      \addplot[blue, mark=star]  table[x index=2, y index=1] {6boxes_grad_1e-4.dat};
      \addplot[magenta, mark=cube]  table[x index=2, y index=1] {6boxes_grad_1e-3.dat};
      \legend{no noise, $\sigma = 10^{-5}$, $\sigma = 10^{-4}$, $\sigma = 10^{-3}$};

      \node[right] at (axis cs:6.5e-05, 2.8e+01) {$\alpha = 10^{-9}$ (over-fitting)};
      \node[above right] at (axis cs:7.5e-05, 6.2e-01) {$\alpha_\text{opt} = 5 \cdot 10^{-5}$ (optimal)};
      \node[left] at (axis cs:3.0e-04, 2.98e-03) {\begin{tabular}{r} $\alpha = 5 \cdot 10^{-3}$ \\ (over-smoothing) \end{tabular}};
    \end{axis}
  \end{tikzpicture}
  \caption{
    L-curves for the reconstruction of the flower state for different noise levels $\sigma$.
  }
  \label{fig:lcurve}
\end{figure}
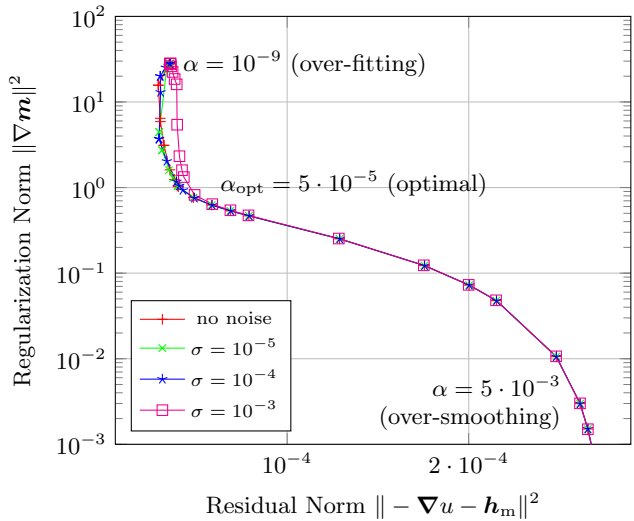

Using the so-called L-curve method \cite{hansen_use_1993}, allows to visualize the trade-off between the reconstruction of the strayfield and the fulfillment of the regularization constraint. Plotting the regularization norm (also called solution norm) $\| T(\mathbf{m}) \|$ over the residual norm $\| -\vec{\nabla} u - \vec{h}_\text{m} \|$ for different regularization parameters $\alpha$, shows an L-shaped curve. The optimal $\alpha_\text{opt}$ can be selected at the corner of the L-curve which means that $\alpha$ is large enough to reduce the regularization norm significantly, but it does not change the residual norm too much. The resulting L-curves for the reconstruction of the flower state for different noise levels are summarized in Fig. \ref{fig:lcurve}. An optimal value $\alpha_\text{opt} \approx 5 \cdot 10^{-5}$ can be found.

The application of the presented method to optimal design problems should now be demonstrated by means of a Halbach cylinder configuration. The goal is to find a magnetization configuration within a cylindrical domain, which creates a homogeneous strayfield inside of the cylinder. The magnetization domain $\Omega_\text{m}$ has an outer radius $r_o = 1.0$, an inner radius $r_i = 0.6$, and a height of $h=2.0$, while a cylindrical measurement domain $\Omega_\text{h}$ with radius $r_m=0.5$ and the same height is used. The magnetization vectors should have constant norm, which suggests using the constant-norm penaltization functional \eqref{eqn:constant_norm_regularization}. The analytic solution for a cylindical Halbach array \cite{halbach_design_1980} can be expressed in cylindrical coordinates as
\begin{align} \label{eqn:halbach}
  \mathbf{m}(\rho, \phi) = \cos(\phi) \; \mathbf{e}_\rho - \sin(\phi) \; \mathbf{e}_\phi
\end{align}
where $\rho$, $\phi$ are the cylindrical coordinates, with the corresponding unit vectors $\mathbf{e}_\rho$, $\mathbf{e}_\phi$.

\begin{figure}
  \centering
  \begin{tikzpicture}
    \node[anchor=south west,inner sep=0] (image) at (0,0) {
      \includegraphics[width=0.7\columnwidth]{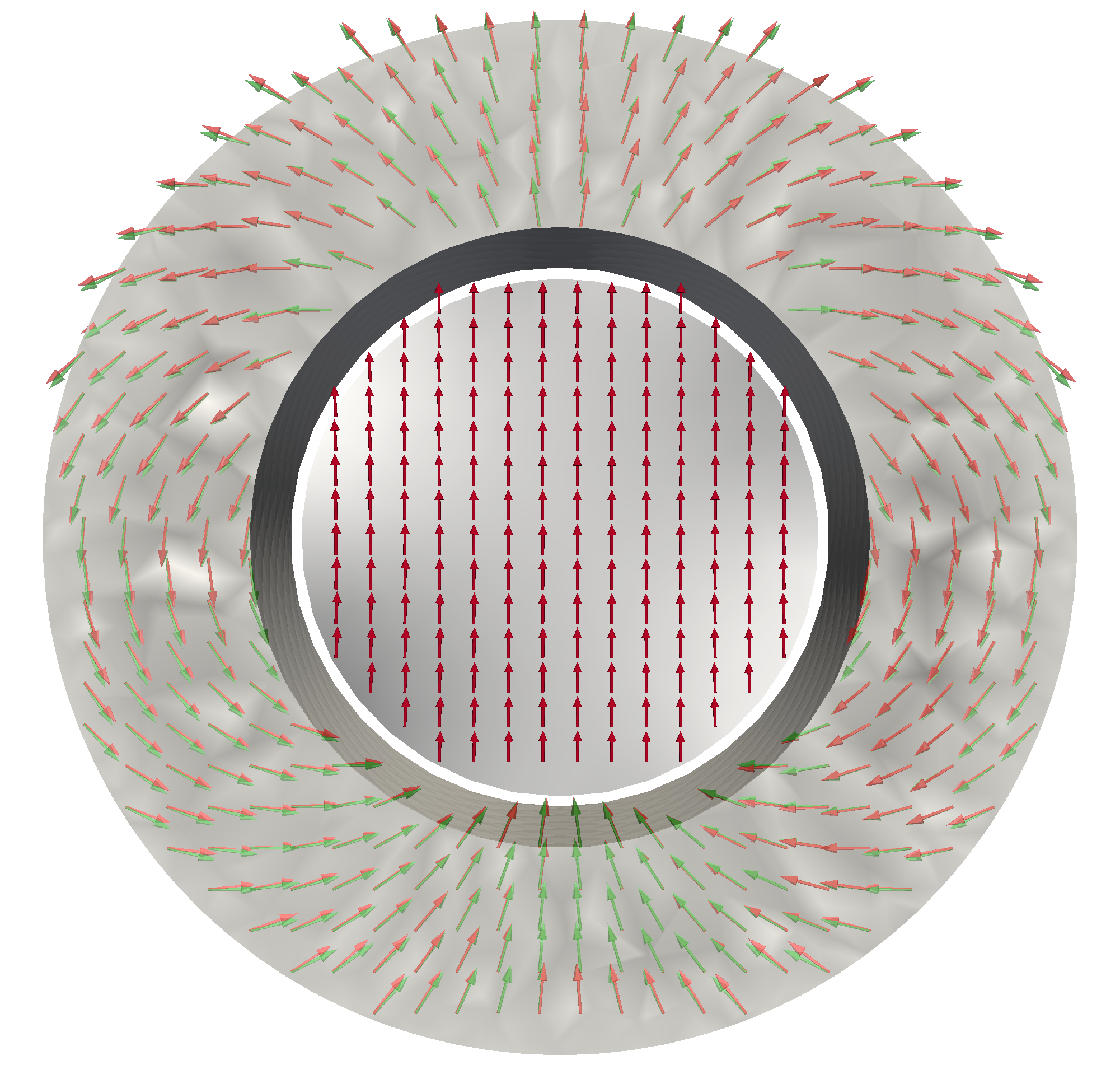}
    };
    \begin{scope}[x={(image.south east)},y={(image.north west)}]
      \draw[fill] (0.72, 0.50) circle[radius=0.005, fill] -- (0.96, 0.70) node[right] {$\Omega_\text{h}$};
      \draw[fill] (0.82, 0.82) circle[radius=0.005, fill] -- (0.89, 0.88) node[right] {$\Omega_\text{m}$};
    \end{scope}
  \end{tikzpicture}
  \caption{
    Optimal design problem of a Halbach cylinder creating a homogeneous strayfield inside of the cylinder. Starting from a homogeneous strayfield the presented algorithm reproduces a Halbach like magnetization configuration within $\Omega_\text{m}$ (red arrows). A constant-norm regularization with $\alpha = 10^4$ is used and shows a nearly perfect match with the analytical solution (green arrows). The resulting strayfield is calculated inside $\Omega_\text{h}$ and shows a nearly homogeneous distribution (red arrows). The relative errors of the magnetization magnitude and the reconstruced strayfield are indicated by the gray-scale, and their maximum amount to $2\%$ and $6\%$, respectively.
  }
  \label{fig:halbach}
\end{figure}

As visualized in Fig. \ref{fig:halbach} there is a nearly perfect match of the reconstructed and the analytical Halbach configuration.

\section{Conclusion}
An efficient algorithm for the solution of inverse problems has been introduced. The use of the Finite Element library FEniCS allows to easily define application specific regularization functionals in a very flexible way. Thus, the implemented algorithm is suitable for a wide range of applications including reverse engineering of magnetic components, design and optimization of magnetic circuits and topology optimization, respectively. Using the hybrid FEM-BEM method proposed by Fredkin-Koehler allows to handle the open-boundary problem accurately and without the need for global mesh including a large airbox. Source identification has been validated by the successful reconstruction of the magnetic flower-state within a unit cube by means of Tikhonov regularization. The selection of a suitable regularization parameter has been demonstrated using the L-curve method. Finally, the application of the method to an optimal design problem has been demonstrated by means of an Halbach cylinder, which is nearly perfectly reproduced.

\section{Acknowledgments}
The authors acknowledge the CD-Laboratory AMSEN (financed by the Austrian Federal Ministry of Economy, Family and Youth, the National Foundation for Research, Technology and Development), the Vienna Science and Technology Fund (WWTF) under Grant No. MA14-044 and the Austrian Science Fund (FWF) under gant No. I2214-N20 for financial support.

\bibliographystyle{ieeetr}
\bibliography{refs}

\end{document}